\documentclass[11pt]{article}
\usepackage{amssymb,amsmath,cite,geometry,graphicx,url,mathrsfs}
\catcode`\@=11 

\@addtoreset{equation}{section}
\def\theequation{\arabic{section}.\arabic{equation}}
     
\catcode`\@=11
\def\thesection{\arabic{section}}

\def\appendix{\setcounter{section}{0}
        \def\thesection{Appendix.}
        \def\theequation{\Alph{section}.\arabic{equation}}}
\def\section{\@startsection{section}{1}{\z@}{3.5ex plus 1ex minus
   .2ex}{2.3ex plus .2ex}{\large\bf}}
     
\long\def\@makefntext#1{\parindent 0cm\noindent
\hbox to 1em{\hss$^{\@thefnmark}$}#1}

\newcommand{\captionfonts}{\small}
\makeatletter  
\long\def\@makecaption#1#2{%
  \vskip\abovecaptionskip
  \sbox\@tempboxa{{\captionfonts #1: #2}}%
  \ifdim \wd\@tempboxa >\hsize
    {\captionfonts #1: #2\par}
  \else
    \hbox to\hsize{\hfil\box\@tempboxa\hfil}%
  \fi
  \vskip\belowcaptionskip}
\makeatother   

\def\nord#1{\mathop{:}\nolimits\!#1\!\mathop{:}\nolimits}
\DeclareMathOperator{\sgn}{sgn}

\begin{document}
\begin{titlepage}
\vspace{.5in}
\begin{flushright}
September 2022\\  
revised March 2023\\
 \end{flushright}
\vspace{.5in}
\begin{center}
{\Large\bf
 Spacetime foam: a review}\\  
\vspace{.4in}
{S.~C{\sc arlip}\footnote{\it email: carlip@physics.ucdavis.edu}\\
       {\small\it Department of Physics}\\
       {\small\it University of California}\\
       {\small\it Davis, CA 95616}\\{\small\it USA}}
\end{center}

\vspace{.5in}
\begin{center}
{\large\bf Abstract}
\end{center}
\begin{center}
\begin{minipage}{4.7in}
{\small
More than 65 years ago, John Wheeler suggested that quantum uncertainties
of the metric would be of order one at the Planck scale, leading to large
fluctuations in spacetime geometry and topology, which he termed ``spacetime foam.''  
In this review I discuss various attempts to implement this idea and to test it, both
theoretically and, to a lesser extent, observationally.
 }
\end{minipage}
\end{center}
\end{titlepage}
\addtocounter{footnote}{-1}

\section{ Wheeler's foam \label{Whee}}

The search for a quantum theory of gravity is at least in part an effort to understand
the very small scale structure of spacetime.  From the earliest days of the search, there
were indications that the Planck length
\begin{align}
\ell_p = \sqrt{\frac{\hbar G}{c^3}}
\label{a1}
\end{align}
determined what ``very small scale'' meant \cite{Bronstein}.  But exactly how quantum
gravity manifested itself at the scale remained---and still remains---a mystery.

In the mid-1950s, Wheeler made one of the first concrete conjectures about that small scale
structure \cite{Wheeler1,Wheeler2}.  He argued that quantum fluctuations of the metric
should be of order one at the Planck scale, leading to wild and rapidly varying fluctuations
in spacetime geometry and topology, which he called ``spacetime foam.''  Later, he would
give a popular description \cite{Wheeler3}:
\begin{quote}
 To the transatlantic passenger flying above it, the  ocean appears smooth. When the traveler's 
 plane has dropped in altitude to a hundred meters above the  surface he sees the waves. 
 When he is in the lifeboat he sees the foam forming and breaking, breaking and forming. 
 Space, smooth at the everyday level, and smooth even at distances comparable to atoms, 
 atomic nuclei and elementary particles, is predicted at still smaller distances to show a foamlike 
 structure.
 \end{quote}
 
 Wheeler gave several arguments for the existence of spacetime foam.  Perhaps the simplest
 was this \cite{MTW}.  Consider a region of space, say a cube, with volume $L^3$.  The smallest quantum
 gravitational excitation that will fit within that cube is a graviton of wavelength $L$, and therefore
 energy $\sim\hbar c/L$, giving an energy density
 \begin{align}
 \rho \sim \frac{\hbar c}{L^4}  .
 \label{a2}
 \end{align}
 On the other hand, we know from classical general relativity that a small metric fluctuation
 $\delta g$ within this cube will have an effective energy density \cite{Isaacson}
 \begin{align}
 \rho \sim \frac{c^4}{G}\left(\delta g/L\right)^2  .
 \label{a3}
 \end{align}
Equating these two expressions, we obtain
\begin{align}
\delta g \sim\frac{\ell_p}{L}  ,
\label{a4}
\end{align}
confirming that metric fluctuations become order one at the Planck scale.

This argument is far from conclusive, of course.  For one thing, it requires an extrapolation of
the energy density (\ref{a3}), derived for small fluctuations of the metric, to order one 
fluctuations.  But a host of other thought experiments \cite{Hossenfelder}, many of them
more recent, point to the Planck length as the scale at which quantum fluctuations become
large.  There are, of course, loopholes: if gravity is asymptotically free, for instance, interactions
may become weak near the Planck scale.\footnote{For the more subtle case of asymptotic safety, 
see \S3.5 of \cite{Hossenfelder}.}  A full answer would require a much more complete quantum 
theory of gravity, but the picture of spacetime foam is certainly worth exploring.

In this paper I will give an incomplete review of recent work on spacetime foam, emphasizing
theoretical aspects but also briefly discussing proposed observational and experimental tests.
I emphasize the word incomplete: ``spacetime foam'' is a slogan attractive enough to have
inspired a great deal of work, while vague enough to encompass a huge range of possible 
quantum gravitational effects.  I will try to emphasize those ideas that are closest
to Wheeler's original vision, but I will undoubtedly miss some.

Note also that for now, questions about spacetime foam, including the most basic one---whether 
it exists---will have answers that depend on one's approach to quantum gravity.  We have a
wide range of research programs aimed at quantizing gravity \cite{CarlipQG,Kiefer}, and
although some  are more popular than others, it's fair to say that none has yet given us a
complete theory.  In what follows, I will try to be careful to specify assumptions, but the
reader should keep in mind that any claim about spacetime foam is provisional.  The reader 
should also keep in mind that the term ``foam'' is a metaphor, not a literal description, and
that not everything that has to do with spacetime and can be thought of as ``foam'' is necessarily
Wheeler's spacetime foam.

\section{Constructing foamy spacetimes \label{construct}}\setcounter{footnote}{0}

Before proceeding with the physics, it is worth taking some time to look at the kinds of geometries
we should include in ``spacetime foam.''  The term ``quantum fluctuations'' is, of course, not very 
precise; at various times, Wheeler specified this to mean either ``configurations that contribute 
most to the sum over histories'' \cite{Wheeler2} or, from the canonical point of view, ``configurations 
that occur [given a wave function] with practically the same probability'' as the most probable
``classical'' configuration \cite{MTW}.  Certainly, then, we should not restrict ourselves to
solutions of the classical field equations.

On the other hand, in many approaches to quantum gravity we \emph{should} restrict ourselves
to configurations that satisfy the constraints.  Recall that in the ADM formalism, the Einstein-Hilbert
action is  
\begin{align}
 I_{\textnormal{\tiny grav}} = \int dt\int_\Sigma d^3x\left(\pi^{ij}\partial_tq_{ij}
    - N\mathscr{H} - N_i\mathscr{H}^i \right) 
\label{c1}
\end{align}
with
\begin{align}
\mathscr{H}^i &= -2\,{}^{\scriptscriptstyle(3)}\nabla_j\pi^{ij}   \label{c2a} \\
\mathscr{H} &= \frac{2\kappa^2}{\sqrt{q}}\left(\pi^{ij}\pi_{ij} - \frac{1}{2}\pi^2\right) 
    - \frac{1}{2\kappa^2}\sqrt{q}\,\,({}^{\scriptscriptstyle(3)}\!R - 2\Lambda) , \label{c3a}
\end{align}
where $q_{ij}$ is the spatial metric, $\pi^{ij}$ is its conjugate momentum, and $\kappa^2=8\pi G$.
(I use the conventions of \cite{Carlipbook}.)  The lapse function $N$ and the shift vector $N^i$  
appear only as Lagrange multipliers, and in a Lorentzian path integral they will produce delta
functionals $\delta[\mathscr{H}^i]$ and $\delta[\mathscr{H}]$ of the constraints, although the latter 
depends on the range of integration over $N$  \cite{Teitelboim}.  Thus in this formulation,
the only configurations occurring in the path integral are those with $\mathscr{H}^i = \mathscr{H} =0$.
In other approaches, though---for instance, the Euclidean path integral \cite{Hawking} or the lattice 
path sum of causal dynamical triangulations \cite{Loll}---no such delta functionals occur.  In
such approaches, the constraints typically have the same status as the rest of the classical
field equations; configurations that do not obey the constraints may be dynamically suppressed, 
but  they will still appear in the full path integral.  For now, we do not know which is the correct 
choice.

There is also no consensus over whether fluctuations of topology are possible.  Classically,
the spatial topology of a Lorentzian manifold can change only if the spacetime admits closed timelike
curves or has a singular metric \cite{Geroch,Sorkin}, though the singularity can be quite mild \cite{Horowitz};
see \cite{Borde} for a more thorough discussion.  If one additionally imposes the classical field
equations, spatial topology change requires a violation of a positive energy condition \cite{Tipler}, and
local regions of nontrivial topology evolve to singularities \cite{Gannona,Lee}.  As we will see in  
\S\ref{passive}, though, vacuum fluctuations frequently carry negative energy, so this may not be a 
problem for the quantum theory.  

Not surprisingly, different approaches to quantum gravity suggest different possibilities for allowed topologies.  
At one extreme, one might include all four-manifolds \cite{Wheeler2}, possibly adding pseudomanifolds 
\cite{Hartleb} or conifolds \cite{Schleich,Schleichb} and perhaps including exotic differential structures 
\cite{Brans,Assel,Duston}.  At the opposite extreme, it can be argued that canonical quantization requires a 
fixed topology \cite{AshMar}.  Between these extremes, one might choose to restrict to topologies of the
form $\mathbb{R}\times\Sigma^3$ with an arbitrary three-manifold $\Sigma^3$ \cite{Horowitz}, or to simply 
connected four-manifolds \cite{Hawking}, or perhaps to manifolds satisfying certain boundary conditions 
\cite{Farey,Witten_bc}.  In at least one instance, a discrete causal model of two-dimensional quantum gravity,  
one can choose to sum over topologies or not, with different results \cite{Westra,Ambjornx}.

If we choose to allow all topologies, an added complication appears: four-manifolds are not classifiable,
in the sense that there is no algorithm that can determine whether two arbitrary four-manifolds are 
homeomorphic \cite{classify,VanMeter}.  It is not clear that this is an insurmountable problem, since it may 
still be possible to compute physical observables to an arbitrarily good approximation \cite{GerochHartle}, 
but it is at least a concern.  Simply connected manifolds are under better control \cite{Friedl}, providing
a practical impetus to focus on such topologies \cite{Hawking}.  Note that three-manifolds \emph{are} 
classifiable \cite{Besson}, although this does not necessarily imply that product manifolds 
$\mathbb{R}\times\Sigma^3$ are.

A further complication comes from the fact that for four-manifolds, homeomorphism does not imply
diffeomorphism.  That is, two four-manifolds $M$ and $N$ may be homeomorphic---equivalent under a
continuous function---but not diffeomorphic.  In such a case, $M$ and $N$ are said to have the same
topology but different ``smooth structures'' or ``differential structures.''  In four dimensions, ``exotic'' smooth 
structures are not uncommon: even $\mathbb{R}^4$ has infinitely many \cite{Gompf}, as does 
$\mathbb{R}\times\Sigma^3$ for any compact three-manifold $\Sigma^3$ \cite{Bizaca}.  In fact, for any integer 
$k$ one can find a four-manifold with $k$ smooth structures that admit (different) Einstein metrics and 
infinitely many other smooth structures that admit no Einstein metric \cite{Braungardt}.  It has even been
suggested that this proliferation of smooth structures in four dimensions might explain the dimension of
spacetime: perhaps all dimensions should be included, but the four-dimensional structures ``swamp the 
path integral'' \cite{Freund}.

The question of what manifolds to allow in a description of spacetime foam is thus complicated, and
we seem rather far from a consensus, or even an understanding of how a decision might be reached.
Still, explicit constructions of foamy spacetimes have been carried out in at least two different 
contexts, the Euclidean path integral (gravitational instantons) and the canonical formalism (foamy initial data). 

In the remainder of this section, I will focus on vacuum solutions, that is, spacetimes for which the stress-energy
tensor vanishes, except perhaps for a cosmological constant.  Omitting matter is a physically reasonable assumption.
We are interested in geometry at the Planck scale, and except perhaps very near a singularity, the effects of matter 
at such short distances are
negligible; even a neutron star has a density of only $10^{-79}$ in Planck units.  A possible exception comes 
from quantum fluctuations of the vacuum stress-energy tensor, which could be large at very small scales.
These are discussed in \S\ref{passive}, but a better understanding of their backreaction on the geometry
would be useful.

Note, though, that we also expect the effective gravitational action to acquire higher order corrections
from quantum effects.  In particular, vacuum fluctuations of matter fields will induce quadratic terms
in the curvature \cite{DeWittx}, leading to an effective action
\begin{align}
I = -\frac{1}{2\kappa^2} \int\!d^4x\,\sqrt{|g|}\left(R + 2\Lambda + \alpha C_{abcd}C^{abcd} 
     + \beta R^2\right) ,
\label{c3b}
\end{align}
where $C_{abcd}$ is the Weyl tensor.   The coefficients
$\alpha$ and $\beta$ are expected to be of order $\ell_p{}^2$, and the corrections are likely to be
negligible at ordinary scales, but they may become important near the Planck scale.

Fortunately, in four dimensions the extrema of the usual Einstein-Hilbert action---that is, vacuum 
solutions of the Einstein field equations with an arbitrary cosmological constant---are also extrema of the  
quadratic action (\ref{c3b}) for arbitrary values of $\alpha$ and $\beta$ \cite{Anderson,Pravda}.  But
(\ref{c3b}) also has other extrema, whose role is not well understood.  If one adds higher order curvature
terms, things become more complicated; Einstein spaces may no longer be exact solutions, 
although one might expect the corrections to be small.  There is some evidence that perturbative quantum
gravity is ``asymptotically safe'' and that corrections to the action beyond quadratic order become unimportant 
at very short distances \cite{Gubitosi,Falls}, but this is far from being proven.
 
\subsection{Gravitational Instantons \label{instantons}}

In the Euclidean path integral approach to quantum gravity, described below in \S\ref{pathint}, the saddle
point approximation is determined by four-manifolds $M$ with Riemannian (that is, positive definite) metrics
that obey the classical field equations 
\begin{align}
R_{\mu\nu} = \Lambda g_{\mu\nu}  .
\label{c4}
\end{align}
To a mathematician, these are Einstein manifolds  with Einstein metrics; to a physicist, they are sometimes  
referred to as gravitational instantons \cite{Hawkingd,Hawkingc,GibbonsHawking,GibbonsHawkingb}.  
As noted above, such metrics continue to be extrema even if quadratic corrections appear in the action, although 
as noted above, there may then be other extrema as well \cite{Anderson}.  

To isolate the dependence on $\Lambda$, it is convenient to
rescale the metric to
\begin{align}
g_{\mu\nu} = \frac{3}{|\Lambda|}{\tilde g}_{\mu\nu} ,
\label{c5}
\end{align}
where $\tilde g$ now solves (\ref{c4}) with $\Lambda=\pm3$, and has a scalar curvature of $\pm12$.
The ``normalized volume''
\begin{align}
{\tilde v}(M) = \int_M\!d^4x\,\sqrt{{\tilde g}}
\label{cx}
\end{align}
is a geometric quantity, but is in some sense topological: different topological four-manifolds have 
different spectra of normalized volumes, and ${\tilde v}(M)$ is believed to be related to Gromov's
invariant ``minimal volume'' \cite{Gromov,CarlipEucl}.  We will see in \S\ref{pathint} that a central quantity
in the Euclidean path integral is the ``density of topologies'' $\rho({\tilde v})$---the number of metrics, typically 
coming from topologically distinct four-manifolds, with normalized volume $\tilde v$.

The question, then, is which four-manifolds admit Einstein metrics, and with what normalized volumes.
This is a very active area of research in mathematics; see \cite{Besse,Andersonb} for reviews.  Note first
that not every four-manifold admits an Einstein metric at all, with any value of $\Lambda$.  Indeed, even such 
simple manifolds as $T^4\#T^4$, the connected sum of two tori,\footnote{The connected sum $M\#N$ of two
manifolds is formed by cutting a solid ball out of each and then identifying the resulting spherical boundaries.  
To a physicist, this amounts to attaching $M$ and $N$ by a wormhole.} and $S^1\times S^3$ have
no Einstein metrics \cite{Besse}.  One simple obstruction was found by Hitchin \cite{Hitchin} and Thorpe
\cite{Thorpe}: for $M$ to admit an Einstein metric, two topological invariants, the Euler characteristic $\chi$ 
and signature the $\tau$, must obey the inequality
\begin{align}
\chi(M)\ge\frac{3}{2}|\tau(M)|  .
\label{c7}
\end{align}
This is a necessary condition, not a sufficient one; stronger limits have been found by a number of
authors \cite{LeBruna,Sambusetti,LeBrunb,Kotschick}.

For $\Lambda>0$, we have a few explicit examples of compact manifolds admitting Einstein metrics.  The sphere 
$S^4$ has its usual constant curvature metric, which can be shown to be the Einstein metric with the largest
normalized volume $\tilde v$ \cite{Bishop}.  The complex projective plane $\mathbb{CP}^2$ admit an Einstein 
metric, as do connected sums $\mathbb{CP}^2\#\,k{\overline{\mathbb{CP}}}^2$ with $3\le k \le 8$ \cite{Tian}.  
But the known cases are surprisingly scarce, and there are some  rather strong restrictions.  Notably, Myers' 
theorem implies that a complete compact Einstein manifold with $\Lambda>0$ necessarily has a finite 
fundamental group: there are no ``Euclidean wormholes'' with positive cosmological constant \cite{Myers,Besse}.  
A compact four-manifold admitting a spin structure has no Einstein metric metric with  $\Lambda>0$ (and, in
fact, no metric with $R>0$) unless the signature $\tau(M)$ vanishes (see \cite{Besse}, \S6F).  
Similarly, for $\Lambda=0$ a few Einstein manifolds are known, notably the tori and certain complex manifolds 
(K3 surfaces), but explicit examples are again scarce.

For $\Lambda<0$, we understand a much richer variety of Einstein manifolds.  To begin with, 
there are infinitely many hyperbolic four-manifolds, that is, manifolds with constant negative curvature
\begin{align}
R_{\mu\nu\rho\sigma} = \frac{\Lambda}{3}(g_{\mu\rho}g_{\nu\sigma} - g_{\mu\sigma}g_{\nu\rho}) .
\label{c8}
\end{align}
Any such manifold is completely determined by its fundamental group $\pi_1$; it is, in fact, a quotient $\mathbb{H}^4/\pi_1$ 
of hyperbolic space, and the constant negative curvature metric is its unique Einstein metric \cite{Bessonb}.  
The density of topologies $\rho({\tilde v})$ for such manifolds grows superexponentially with normalized
volume, $\rho({\tilde v})\sim e^{a{\tilde v}\ln{\tilde v}}$ \cite{CarlipEucl,Lubotzky,Gelander}.  Another superexponentially 
growing class of Einstein manifolds, this time not of constant curvature, can be obtained from hyperbolic manifolds 
with mild singularities (``cusps'') by filling in the cusps with a procedure known as Dehn surgery \cite{Andersonb,Andersonc}.  
Yet another superexponentially growing class consists of product manifolds $\Sigma_{g_1}\times\Sigma_{g_2}$ of 
surfaces of genus $g_1$ and $g_2$ \cite{CarlipEucl}.

For $\Lambda\ge0$, the known examples of Einstein metrics fit one aspect of Wheeler's definition of spacetime 
foam: they have complicated topologies that may contribute to the path integral in an amount comparable to  
the simplest $S^4$ topology.  But viewing these as Planck scale fluctuations is not so obvious.  The topological structure 
seems more global; in particular, by Myers' theorem, there are no wormhole-like contributions.  Of course, to be fair,
a Euclidean instanton in conventional quantum field theory represents the overall effect of an enormous number of   
individual Lorentzian configurations, so one should perhaps not expect too much microscopic detail.

An obvious 
way to incorporate Planck scale structure would be to consider connected sums of Einstein spaces, that is,
spaces attached by wormholes.  Unfortunately, this fails: the connected sum of two Einstein manifolds need not 
itself admit an Einstein metric.   It has been suggested that multi-wormhole configurations that are ``near  extrema'' 
may be important \cite{Coleman}, but this is controversial (see, e.g., \cite{Polchinski}).  

For the case $\Lambda<0$, some of the known Einstein manifolds look more like Wheeler's foam.  Anderson's Dehn 
surgery construction \cite{Andersonb,Andersonc}, for instance, yields manifolds in which the nontrivial geometry manifests
itself in a discrete set of ``bubbles,'' regions near the tips of would-be cusps.  But these regions are ``far away''
from each other---the initial cusps are infinitely long---rather than occurring at Planck density.

 For some purposes, one can obtain more ``foam-like'' Einstein metrics by dropping the requirement of 
 compactness.  Depending on the physical context, one may then be interested in instantons that are either 
 asymptotically locally Euclidean \cite{GibbonsHawking,Hitchinb} or asymptotically locally flat and 
 periodic in imaginary time \cite{GibbonsHawkingb}.  ``Locally'' here means ``up to an identification by a discrete
 group''; for mathematical details, see \cite{Eguchi}.  One large class of such instantons  consists of metrics with 
 self-dual or anti-self-dual curvature \cite{Eguchib,Bourliot}; Torre has studied the moduli space of such solutions 
\cite{Torrea,Torreb}.   A particular class of self-dual solutions, the  multi-instanton metrics of Gibbons and 
Hawking \cite{GibbonsHawking}, can even be extended to include spaces of infinite topological type 
\cite{Kronheimer,Nergiz,Calderbank}, although there are potentially problems with singularities and
divergences \cite{Rutledge}.
 
\subsection{Initial data \label{initial}}

A different approach to spacetime foam starts with initial data,\footnote{As noted above, quantum corrections
may lead to higher order field equations, for which it is not known whether the initial value problem is well
posed.  Even then, though, the initial data described here can be evolved with the standard second order 
Einstein equations, yielding spacetimes that remain solutions to quadratic gravity.}
 that is, with the specification of a three-manifold $\Sigma^3$ with 
a metric for which the constraints (\ref{c2a})--(\ref{c3a}) vanish.  As noted earlier, if $\Sigma^3$ has localized
nontrivial topology and the stress-energy tensor obeys a positivity condition, the classical evolution of such 
data will lead to singularities \cite{Gannona,Lee}.  But it is not clear that this is a problem for the quantum
theory \cite{Carlipmini}, where it is expected---or at least hoped---that quantum effects will ``resolve''
singularities \cite{Wheeler1}.

Witt has shown that any closed three-manifold $\Sigma^3$, with any nonnegative distribution of energy, 
admits a metric obeying the constraints \cite{Witta}.  The same is true for at least a large class of asymptotically
flat three-manifolds.  The proof is fairly simple, and the construction is quite explicit.  The resulting initial data
are very special, though---in particular, the canonical momentum is proportional to the metric---and it is useful 
to look for more general constructions.

Such generalizations come in two distinct strands.   Morrow-Jones, Witt, and Schleich have carried
out detailed studies of locally spherically symmetric initial data \cite{Witt1,Witt2,Witt3}.   At first sight, the
requirement of spherical symmetry seems very restrictive: it is widely thought that for a vacuum spacetime, 
spherical symmetry leads inevitably to  Schwarzschild or Schwarzschild--(anti) de Sitter spacetimes.  But while 
Birkhoff's theorem implies that a spherically symmetric vacuum spacetime must be \emph{locally} 
isometric to some region of Schwarzschild, local patches can be sewn together to form spacetimes that look
drastically different  \cite{Witt3}.  In particular, one can form connected sums of large classes of three-manifolds 
with such data, including arbitrary quotients of $S^3$, $\mathbb{R}^3$, and $\mathbb{H}^3$, with a construction 
that is almost completely explicit \cite{Witt1}.

A much broader, although less explicit, construction of foamy initial data follows from work by
Chrusciel, Isenberg, and Pollack \cite{Chrusciel,Chruscielb}, who show that one can take a connected sum of
\emph{any} set of three-manifolds with initial data obeying the constraints.  More precisely, let $\Sigma_1$ 
and $\Sigma_2$ be two three-manifolds, with generic\footnote{The proofs fail if the initial data is too
symmetric, in the sense that it evolves to a four-manifold with a local Killing vector.  This is complementary
to the results of Morrow-Jones et al., who require symmetry.} initial data $(g_1,\pi_1)$ and $(g_2,\pi_2)$ that 
satisfies the constraints, again with suitable positive energy conditions.  Pick open sets $U_1\subset\Sigma_1$ 
and $U_2\subset\Sigma_2$.  Choose points $p_1\in U_1$ and $p_2\in U_2$, cut geodesic balls $B_1$ and 
$B_2$ of arbitrarily small radius around each, and join the boundaries.  Then the connected sum 
$\Sigma_1\#\Sigma_2$ admits initial data that exactly coincides with the original data outside $U_1\cup U_2$ 
and is close to the original data, in a suitable norm, inside $U_1\cup U_2$ but outside $B_1\cup B_2$.  Note
that $\Sigma_1$ and $\Sigma_2$ can be topologically complex, but they may also be three-spheres: the
connected sum may join regions of different topology, but may also join fluctuating geometries with no extra
topological structure.  Thus for the initial three-manifold, this construction looks very much like Wheeler's
picture of spacetime foam.

It is not known how much of the space of initial data is covered by this  Chrusciel--Isenberg--Pollack construction.
At the topological level, all three-manifolds have a decomposition as connected sums, unique for orientable
manifolds and almost unique for nonorientable ones \cite{Milnor,Giulini}.  At the geometrical level, though,
the structure of the ``necks'' connecting components is rather special.  The construction should perhaps be
viewed more as an existence proof for ``foamy'' three-geometries that satisfy the constraints rather than a
comprehensive description of such geometries.

\subsection{Other settings \label{other}}

While most work on spacetime foam has focused on instantons or initial data, there has been some
progress in a few other areas.  These include
\begin{itemize} 
\item{\bf Manifolds with boundary}:\\[1ex]
The emphasis in \S\ref{instantons} was on manifolds without boundaries, except perhaps at infinity.  
For many purposes, though, the physically relevant saddle points are manifolds with boundaries (see 
\S\ref{pathint}).   A compact Einstein manifold with a reflection symmetry can be split in half to form an Einstein 
manifold with a single boundary with vanishing extrinsic curvature, called a ``real tunneling geometry'' 
\cite{GibbonsHartle}, which naturally extrapolates between a metric with Riemannian signature and one 
with Lorentzian signature.  The condition of vanishing extrinsic curvature is rather strong, but can be 
understood as part of the requirement for an extremal action \cite{Carliptunnel}.  

As in the compact case, 
only a few examples of Einstein manifolds with boundary are known for $\Lambda\ge0$, but large classes 
are known for  $\Lambda<0$.  The ``density of topologies'' $\rho({\tilde v},h)$ now depends on both the 
normalized volume and the induced metric $h$ on the boundary, and for a boundary with a single component, 
$\rho({\tilde v},h)$ has sharp peaks at certain symmetric values of the boundary metric \cite{Carlippeaks}.
A sum over topologies has also recently been considered in a two-dimensional single boundary model 
\cite{Ambjorn}, using the lattice approach known as causal dynamical triangulations (see \S\ref{lattice}).

\item{\bf Lorentzian saddle points}:\\[1ex]
The Riemannian Einstein metrics of \S\ref{instantons} occur as saddle points in a particular formulation
of the gravitational path integral.  In other approaches, saddle points may still satisfy the vacuum field
equations (\ref{c4}), but with Lorentzian signature metrics.  It could be that complex saddle points are also
relevant, especially if the path integral is viewed as a contour deformation; I will return to this briefly
in \S\ref{pathint}.   

Of course, solving (\ref{c4}) 
for Lorentzian metrics is equivalent to finding the vacuum solutions of classical general relativity, something 
I will not try to summarize here.  But it is worth recalling the topological restrictions described at the beginning 
of \S\ref{construct}.  In particular, if one requires stable causality---a slightly stronger version of the condition
that there be no closed timelike curves---then local fluctuations in topology are forbidden \cite{Visser}, 
although a fixed \emph{spatial} topology may be very complex \cite{Visserb}.

\item{\bf Simplicial manifolds}:\\[1ex]
One way to control a sum over metrics is to replace smooth manifolds by simplicial manifolds, either as a 
fundamentally discrete approach to quantum gravity or as an approximation to the continuum.   Note that this 
change of setting may change the physics: there are four-manifold topologies that admit no triangulation, and
therefore would be excluded from the path integral in a simplicial approach (see the discussion in
\cite{Schleichb}).  

A fair amount is known about how to count simplicial manifolds, although the results are rather complicated
and many questions remain open \cite{Carfora,Carforab}.   In particular, if the topology is allowed to vary,  
the number of simplicial manifolds grows factorially with the number of simplices \cite{Chapuy}.  The situation
may be different if the topology is fixed.  In two
dimensions, the number of triangulations of the sphere grown only exponentially \cite{Chapuy}, and it
has been conjectured that this is true in higher dimensions as well \cite{Durhuus,Gromovb}, but the question 
remains open.   

\item{\bf Lower dimensions}:\\[1ex]
In fewer than four spacetime dimensions, the characterization of topology and the identification
of Einstein metrics becomes much simpler.  The topology of three-manifolds is now completely classified,
thanks to Perelman's proof of Thurston's geometrization conjecture \cite{Besson}.  Einstein
manifolds in three dimensions are further restricted: they necessarily have constant curvature (that is, they 
satisfy (\ref{c8})), and can always be represented as quotients
\begin{align}
M = {\widetilde M}/\Gamma ,
\label{c9}
\end{align}
where $\widetilde M$ is the round three-sphere $S^3$ (for $\Lambda>0$), flat $\mathbb{R}^3$ (for 
$\Lambda=0$), or hyperbolic three-space $\mathbb{H}^3$ (for $\Lambda<0$), and  where $\Gamma$ is a 
discrete group of isometries of $\widetilde M$.  

For $\Lambda>0$, the spaces of constant curvature were classified by 
Seifert and Threlfall in 1930 \cite{Seifert}.  They fall into a few distinct classes, determined by the type of isometry
group $\Gamma$; some recent references are \cite{Wolf,Wittx}.  For $\Lambda=0$, there are a total of
ten flat closed three-manifolds; the three-torus is the simplest, but the sides of a cube can also be identified 
with various twists \cite{Wolf,Thurston,Conway}.  For $\Lambda<0$, there are an infinite number of compact hyperbolic 
three-manifolds, whose properties have been extensively, although not completely, characterized 
\cite{Thurston,Thurstonb,Ratcliffe}.

The two dimensional case is even simpler.  A compact, orientable two-manifold---that is, a closed
surface---has a topology determined by its genus $g$, the ``number of handles.''  A constant
curvature metric necessarily has positive curvature on a sphere ($g=0$), vanishing curvature
on a torus ($g=1$), and negative curvature on a surface of genus $g>1$.  The space of such 
metrics, the moduli space, has been the subject of a great deal of study, both in mathematics
and in physics; see, for instance, \cite{DHoker,Harvey}.  In some cases it is useful to generalize
to noncompact and nonorientable surfaces, but this is also quite straightforward.

\end{itemize}

\subsection{A caveat on divergent sums \label{caveat}}

One more mathematical issue will be important later.  On occasion, we will find that certain sums over geometries 
or topologies diverge, and we will have to make sense of this behavior.  Of course, many routine divergences 
can be eliminated with, for example, zeta function regularization \cite{zeta} or Borel summation \cite{Borel}.  
Indeed, there are phenomena such as the Casimir effect in which an appropriately regulated sum of 
vacuum fluctuations gives an experimentally verifiable prediction \cite{Romeo}.  But we will sometimes 
find sums for which no obvious regularization exists.

It is important not to overinterpret such divergences.  To see the pitfall, consider two series
\begin{align}
S_1 = \sum_{n=1}^\infty \frac{(-1)^{n+1}}{n}, \quad 
S_2 =  \sum_{n\ \mathrm{even}} \frac{(-1)^{n+1}}{n}  .
\label{c10}
\end{align}
$S_1$ is conditionally convergent; while its value can be changed by systematically rearranging the
terms, as written $S_1=\ln 2$.  $S_2$ is a subseries of $S_1$, but is divergent; even with zeta function 
regularization, $S_2= -\frac{1}{2}\zeta(1)$, which is infinite.  If we were just given $S_2$, the temptation
would be to say that we are summing too many terms, and that perhaps we should cut off the sum at a finite $n$.
But it could be that we were summing too \emph{few} terms, and that the addition of more geometries or 
topologies might give us a physically sensible answer.

\section{Quantum fluctuations of the stress-energy tensor \label{passive}}\setcounter{footnote}{0}

We now return to our central subject, the small scale structure of quantum gravity.
Broadly speaking, quantum fluctuations of the metric come in two forms \cite{Ford}: ``active'' 
fluctuations, which come from the quantum nature of the metric itself, and ``passive'' fluctuations,
the metric's response to quantum fluctuations of the stress-energy tensor.  Passive fluctuations
have been studied for decades \cite{Blok}, notably through a model known as stochastic 
gravity \cite{Hu}.  Until recently, though, most of this work focused on small fluctuations that 
could be treated at low order in perturbation theory.  Such fluctuations can have interesting 
physical effects---see, for instance, \cite{Fordb}---but they are not really spacetime foam.
Over the past few years, however, a good deal of progress has been made in understanding 
high order stress-energy correlators and large fluctuations, leading to something much closer to
Wheeler's original conception \cite{Fewstera,Fewsterb,Fewsterc,Fewsterd}.

The expectation value $\langle T_{\mu\nu}\rangle$ of the stress-energy diverges, and must be
renormalized.  There are a number of ways of doing this; see, for example, \cite{Hack}.
Even after renormalization, though, higher order correlators
\begin{align}
G_{\mu_1\nu_1\mu_2\nu_2\mu_3\nu_3\dots}(x_1,x_2,x_3,\dots)
  = \langle\psi| \nord{T_{\mu_1\nu_1}(x_1)}\nord{T_{\mu_2\nu_2}(x_2)}\nord{T_{\mu_3\nu_3}(x_3)}\dots|\psi\rangle
\label{b2}
\end{align}
will still have divergences, which will in general be state-dependent \cite{Fordb}.  These remaining 
divergences can be eliminated by integrating $\nord{T_{\mu\nu}}$  against a sampling (or
``smearing'') function \cite{FordRoman},
\begin{align}
{T}_{\mu\nu}[w]  = \int\!d^4x\,w_\tau(x)\nord{T_{\mu\nu}(x)}  ,
\label{b3}
\end{align}
where $w_\tau(x)$ is a sharply peaked function with characteristic width $\tau$.  The precise 
values of correlators will depend on the choice of smearing, but some key qualitative features are 
largely independent of $w$.

Now pick a component of the stress-energy tensor, say the energy density $\rho=T_{tt}$, and consider
the smeared vacuum correlators
\begin{align}
\mathcal{G}_n[w] = \langle 0|(\rho[w])^n|0\rangle  .
\label{b4}
\end{align}
\begin{center}
\begin{figure}
\begin{center}
\includegraphics[height=2.3in]{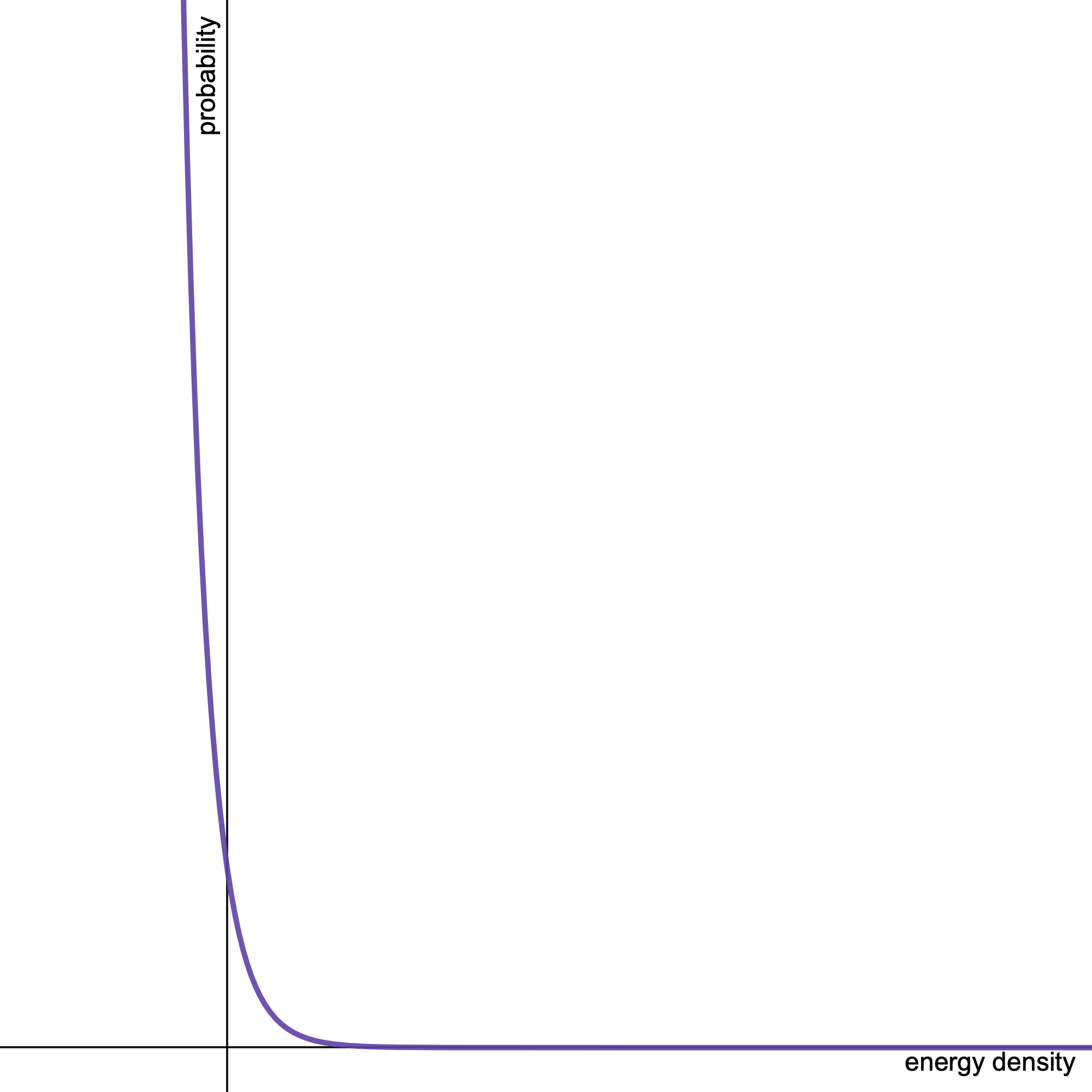}
\end{center}
\caption{Typical distribution of quantum fluctuations of the energy density.  In two dimensions (shown here)
the positive tail decays exponentially, but in four dimensions the decay is slower. 
 \label{fig1}}
\end{figure}
\end{center}
We can ask whether there is any classical probability distribution whose moments reproduce the $\mathcal{G}_n$,
that is, a probability distribution for quantum fluctuations of $\rho[w]$ around its vacuum value of zero.  For 
a conformal field theory in two spacetime dimensions, the answer can be calculated exactly \cite{Fewstera},
and is largely independent of the smearing function \cite{Hollands}.  In terms of the dimensionless variable
$x=\rho\tau^2$,
\begin{align}
P(x) = \theta(x+x_0)\frac{\pi^{c/12}(x+x_0)^{c/12-1}}{\Gamma(c/12)}e^{-\pi(x+x_0)}
\label{b5}
\end{align}
where $c$ is the central charge, $\theta$ is a step function, and  $x_0 = c/{12\pi}$.  This shifted
Gamma distribution, shown in figure \ref{fig1}, has several key features:
\begin{enumerate}
\item Most vacuum fluctuations---for the $c=1$ case,  84\%---are negative.
\item There is, however, a strict lower bound to the energy of a vacuum fluctuation, determined by quantum 
energy inequalities \cite{Fordc}, with a numerical value determined by the smearing function.
\item There is an infinite positive tail.
\end{enumerate}
For the more realistic four-dimensional case, exact results are not available, but there are now good
estimates and bounds \cite{Fewsterb,Fewsterc,Fewsterd}.  The results are qualitatively very similar,
although more sensitive to the choice of smearing function.  In particular, there is still a strict lower bound
and an infinite positive tail, which now  decays slower than exponentially, typically as $e^{-x^{1/3}}$.  Thus
most fluctuations are negative, but there is a ``fat tail'' of very high positive energy fluctuations.

What does this mean for spacetime geometry?  With our choice of renormalization, the vacuum
expectation value of the stress-energy tensor is zero (up to the possible higher order curvature 
corrections discussed briefly in \S\ref{construct}), but the fluctuations can be very large.  
We can treat these as a stochastic source term for the Einstein equations, a procedure that has 
been shown to be a particular next order extension of semiclassical gravity \cite{Hu}.
Of course, the distribution of fluctuations obtained from (\ref{b3}) depends on a smearing length;
this might most naturally taken to be the Planck length, but one can explore the effects of
changing the scale.

In two dimensions, there have been studies of the effect of 
fluctuations of the form (\ref{b5}) on null \cite{Pitelli1} and timelike \cite{Pitelli2} geodesics.  In both cases,
Planck scale fluctuations lead to the convergence of geodesics and the collapse of light cones at scales
of about a hundred Planck lengths.  This behavior can be seen as an example of the phenomenon known
in statistics as ``gambler's ruin'' \cite{ruin}: although most fluctuations have negative energies and defocus 
geodesics, the rare very large fluctuations on the positive tail ultimately dominate.  There has been less
work on the four dimensional case, but a similar behavior appears to occur \cite{Carlipfluc}.  Steps toward a 
more rigorous quantum field theoretic treatment can be found in \cite{Drago}.

These results should probably not be taken too literally: they neglect the correlations between successive 
fluctuations, and do not account for the quantum nature of the gravitational field itself.  But they lend at least 
some support for Wheeler's picture of ``a mechanism which all the time is continually bringing into being black 
holes and wiping them out\dots'' \cite{Wheeler3}.

\section{Quantization}\setcounter{footnote}{0}

While vacuum fluctuations of matter certainly affect the small scale structure of spacetime, the real
motivation for quantum foam comes from quantum gravity.  Unfortunately, we do not yet have a 
complete quantum theory of gravity.  We have a number of promising research programs, and there
are some features that are likely to preserved in any eventual quantum theory of gravity: something
like the Wheeler-DeWitt equation \cite{DeWitt1} is likely to hold, and it is plausible that some form
of the ``holographic'' bulk--boundary duality \cite{Bousso} will survive.  For now, though, different
approaches to quantum gravity may give very different answers to questions about spacetime foam.

In what follows, I will concentrate on the approaches  in which spacetime foam has been studied most 
extensively---the Euclidean path integral, lattice methods, canonical quantization, and lower dimensional
models---though I'll say a bit about some other areas of research.  I will \emph{not} talk about one of
the most interesting recent developments, the possible relationship between wormholes, baby
universes, and the problem of unitarity of black hole evaporation \cite{Marolf}; at this writing, the
matter is too much in flux for me to include in a review article.

Of course, as noted earlier, ``spacetime foam'' is not a terribly well-defined notion.  To be a bit
more precise, I will focus on two indicators, both suggested by Wheeler.  For path integral approaches,
I will ask whether there are large contributions from configurations that are substantially different,
either in topology or in geometry, from the expected classical spacetime geometry.  To some extent, 
this is a matter of choice: we don't yet have an overarching principle to tell us what topologies
ought to be included, but we can ask, for instance, whether sums over topologies make mathematical
sense, and whether transition amplitudes from an initial state to a ``foamy'' final state are identically zero.  
For canonical approaches, I will ask whether the wave function gives high probabilities for ``fluctuations'' 
that are substantially different from the expected classical spacetime geometry near the Planck scale.  
Again, the answer will depend on the particular approach to quantization, but the question, at least, is 
coherent.

\subsection{Path integrals \label{pathint}}

In path integral approaches to quantum gravity, the fundamental quantities take the form
\begin{align}
Z[\Lambda,g_{{\scriptscriptstyle\partial M}}] = \sum_M\int [dg]\exp\{iI[g]\}
\label{d1}
\end{align}
where $I[g]$ is the Einstein-Hilbert action, perhaps analytically continued and perhaps including higher 
order corrections.  The sum in (\ref{d1}) is over some collection of four-manifolds; as discussed in 
\S\ref{construct}, the choice of what to sum over is part of a decision of how to treat spacetime foam.  

The physical meaning of the path integral depends on the characteristics of the manifolds $M$, and in 
particular the nature of their boundaries.  Restricted to compact manifolds without boundary, the path
integral yields a partition function for a ``volume canonical ensemble'' \cite{Hawking,Hawkingc}.  For 
configurations with a fixed periodicity in imaginary time, this is essentially an ordinary thermal partition 
function.  Restricted to manifolds with a single boundary, the path integral gives a functional of the
boundary metric (or perhaps the conformal geometry and mean curvature), which
can be interpreted as a Hartle-Hawking \cite{HartleHawking}  or Vilenkin \cite{Vilenkin} amplitude for 
the creation of a universe from ``nothing.''  Applied to manifolds with more than one boundary, the same
path integral can be understood as a transition amplitude between spatial geometries.

Further details depend on one's choice of which metrics to include in the path integral.  Spacetime foam
has been studied most extensively in the context of the Euclidean path integral 
\cite{Hawking,Hawkingc,Hawkingb,Coleman},
in which the integral (\ref{d1}) is taken over Riemannian  (positive definite) metrics, and the
appropriately analytically continued path integral is
\begin{align}
Z[\Lambda] = \sum_M\int [dg]\exp\{-I_E[g]/\hbar\} \quad\hbox{with}\ \ 
I_E[g] = -\frac{1}{2\kappa^2}\int_M\!d^4x\,\sqrt{g}(R-2\Lambda)  .
\label{d2}
\end{align}
This choice has been justified as a contour deformation of the Lorentzian path integral, though there
is controversy over the correct choice of saddle points in such a deformation \cite{Turok}.  But it can
alternatively simply be postulated as a definition of the quantum gravity path integral, one that
automatically yields wave functions that satisfy the Wheeler-DeWitt equation \cite{Halliwell}.

The saddle points of the Euclidean action $I_E$ are the Einstein metrics
\begin{align}
R_{\mu\nu} = \Lambda g_{\mu\nu}  
\label{d3}
\end{align}
discussed in \S\ref{instantons}.  As noted there, these remain saddle points even when quadratic corrections
are added to the action.  With the rescaling (\ref{c5}), the saddle point approximation to the path integral
(\ref{d2}) becomes
\begin{align}
Z[\Lambda] = \sum_{\tilde v} \rho({\tilde v})\exp\left\{\frac{9}{\hbar\kappa^2\Lambda}{\tilde v}\right\} ,
\label{d4}
\end{align}
where the sum is over saddle points, $\tilde v$ is the normalized volume (\ref{cx}), and $\rho({\tilde v})$ is 
the density of topologies with normalized volume $\tilde v$.  In a slightly more sophisticated analysis, one 
can also include in $\rho({\tilde v})$ the one-loop contributions $\Delta_{(M,g)}$  to the path integral, 
which involve Faddeev-Popov determinants from gauge-fixing and Van Vleck-Morette determinants from 
fluctuations around the extrema.  While these are complicated, and not completely calculable, their 
dependence on the normalized volume can be computed from the trace anomaly \cite{Christensen}.  
Crucially, the one loop factor is at most exponential in $\tilde v$:
\begin{align}
\Delta_{(M,g)} \sim e^{b{\tilde v}} \quad\hbox{with $b = \frac{261}{80\pi^2}\ln|\Lambda| $}  .
\label{d5}
\end{align}

To evaluate the sum (\ref{d4}), we would need to know the saddle points, that is, the set of Einstein
manifolds.  We do not.  In \cite{Hawking}, Hawking offers a qualitative discussion, in which he   
conjectures that saddle points with large Euler number $\chi$ must have $\Lambda<0$, and that  
typically $\chi\sim{\tilde v}$.  (Up to constants of order one, Hawking's $f$ in \cite{Hawking} is 
$f = -\sgn\Lambda\sqrt{\tilde v}$.)  As far as I know, it is not known whether this is true; it is related to 
what mathematicians call the ``geography problem,'' the (unsolved) classification problem
of determining the collection of four-manifolds with given Euler number and signature (see, for 
instance, \cite{Fintushel}).

For $\Lambda<0$, though, even without knowing the full set of Einstein manifolds, there are still some things
we can say.  Consider first the partition function,  the sum over compact manifolds.  Recall from 
\S\ref{instantons} that the number of compact Einstein manifolds with $\Lambda<0$ grows faster than
exponentially.  Thus even a partial sum---say, the sum restricted to hyperbolic manifolds, or to manifolds 
obtained from Dehn fillings of cusps---yields
\begin{align}
Z[\Lambda] \sim 
\sum_{\tilde v}\exp\left\{ a{\tilde v}\ln{\tilde v} + \left(b-\frac{9}{\hbar\kappa^2|\Lambda|}\right){\tilde v}\right\} .
\label{d6}
\end{align}
(Here $a$ and $b$ are dimensionless constants coming from mathematical existence theorems
\cite{Lubotzky,Gelander,Andersonc}; $a$ is positive, but their values are not known.)

This sum does not converge.  The ``entropy,'' the growth in the number of saddle points, dominates the 
``action,'' the suppression from the volume dependence of $I_E$.  Whether the expression is Borel 
summable depends on the phases, which arise from the Van Vleck-Morette determinants and are
not known at this time.  The behavior of this sum has been compared to that of a thermal ensemble with 
negative specific heat, with $\tilde v$ serving as ``energy'' and $|\Lambda|$ as ``temperature'' 
\cite{CarlipEucl,Carlip_foam}.  As in such an ensemble, an increase in ``energy'' drives the production of 
new states, whose number
rises so rapidly that the ``temperature'' per state is driven down.  But we should also recall the caveat of 
\S\ref{caveat}: we do not know how to include the full set of saddle points in the sum, and we do not
know relative signs, so it could be that the full path integral behaves better.

We may also consider the sum over manifolds with a single boundary.  This requires us to impose boundary 
conditions; a natural choice is to specify the trace $K$ of the extrinsic curvature and the conformal class 
${\tilde h}_{ij}$ of the induced metric at $\partial M$ \cite{Carliptunnel,Witten_bc}.  This may tame the divergence%
---while the number of compact Einstein manifolds grows superexponentially with $\tilde v$, this need not
imply that the number of Einstein manifolds with fixed $(K,{\tilde h})$ does.  It can be shown, however,
that a build-up of Einstein manifolds that are geometrically close to each other leads to sharp peaks at 
certain highly symmetric boundary geometries \cite{Carlippeaks}.

For $\Lambda>0$, the set of Einstein metrics is even more poorly understood, and much less can be said.
One might worry that the ``action'' term could now lead to a divergence, since the exponent in (\ref{d4}) is now
positive, but as noted in \S\ref{instantons}, the normalized volume takes on its maximum value when $M$
is a four-sphere.  There has been a little work on single boundaries with nontrivial topology (for instance,
\cite{Vargas}), but again, too little is known so far to allow a systematic study.  Similarly, the $\Lambda=0$ case 
is poorly understood, especially since the saddle point action is now zero and the saddle point approximation, 
if it is valid at all, is controlled by the one loop determinants.  

An alternative to the Euclidean path integral is to start with the Lorentzian path integral---that is, the
path integral over metrics with Lorentzian signature---and then look for a contour deformation in
the space of metrics.  For minisuperspace models, this approach has a long history (see, for instance,
\cite{HalliwellLouko}), and it has recently undergone a revival involving more sophisticated methods of
contour deformation \cite{Turok}.  While there has been fairly little work on spacetime foam in this
context, it has recently been suggested, generalizing earlier work by Louko and Sorkin \cite{LS},
 that certain saddle points should be excluded on the basis that they do not allow quantum field theory to 
 be defined \cite{Kontsevich,Witten_cpx,Lehners}.

\subsection{Lattice methods \label{lattice}}

To move beyond the saddle point approximation of the path integral, one possibility is to discretize
spacetime, changing a rather poorly defined ``integral'' into a finite sum that can be evaluated
numerically.  The typical approach is to approximate spacetime as a simplicial manifold, in which
the simplices either have varying edge lengths (``Regge calculus'') or fixed edge lengths but
varying combinatoric configurations  (``dynamical
triangulation'') \cite{Lollrev}.  The hope is that, as in lattice gauge theory, a second order phase 
transition will occur; at such a transition, correlation lengths diverge, essentially erasing the discreteness.   
Concrete calculations use the Regge action \cite{Regge}, the discrete version of the Euclidean 
action (\ref{d2}).

As noted in \S\ref{other}, the number of simplicial manifolds increases faster than exponentially with
volume, while the action $I_E$ grows only exponentially, so we should perhaps not expect
convergence---as in the saddle point approximation with $\Lambda<0$, ``entropy'' would seem 
to dominate ``action.''  Indeed, while there are some disagreements and subtleties \cite{Hamber,Laiho}, 
it is generally believed that traditional lattice approaches have no good continuum limit that resembles
our real spacetime.

 \begin{figure}
\centerline{\includegraphics[height=2.2in]{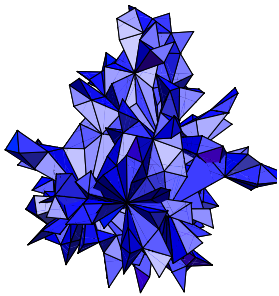}}
\caption{A typical two-dimensional spatial slice in a three-dimensional causal dynamical triangulations
simulation (figure created by Michael Sachs)\label{fig2}}
\end{figure}

An alternative lattice approach shows much better convergence.  In causal dynamical triangulations
\cite{Loll}, a time slicing is chosen in a way that fixes the spatial topology, prohibiting topology
change and at the same time providing an unambiguous direction of time that makes the spacetime 
more ``Lorentzian.''  As in the continuum case \cite{Visser}, causality is accompanied by the 
prohibition of topology change.  This eliminates one version of spacetime foam, the conjectured 
quantum fluctuations in topology, although the initial spatial slice may have an arbitrarily
complicated topology.

But while fluctuations of topology are excluded, large fluctuations of geometry can certainly
occur.  Numerical simulations show that on large scales, the sum over configurations that defines 
the path integral has features that look very much like the expected classical spacetime,
most notably the spatial ``volume profile'' as a function of time.  But a typical configuration in the 
path integral can be very far from a saddle point: see figure \ref{fig2}, for example.
One way to quantify such fluctuations is to look at ``dimensional estimators,'' various measures
of the effective dimension of the space or spacetime \cite{Carlipdim}.  In particular, the spectral
dimension of spacetime---the dimension as measured by a random walk---is, as expected, $d_S=4$
at large scales, but it shrinks to $d_S\approx 2$ at small scales \cite{AmbLolla}, and the spectral dimension
of a slice of constant time is $d_S\approx 1.5$ \cite{AmbLollb}.   

\subsection{Canonical quantization \label{canon}}

General relativity is a totally constrained system: as can be seen from (\ref{c1}), the Hamiltonian, up to possible 
boundary terms, consists entirely of constraints.  This is the starting point for attempts to canonically
quantize the theory.  Typically\footnote{In the alternative reduced phase space quantization, one first
solves the constraints for a set of ``true'' degrees of freedom and then quantizes these.  While this
program can sometimes be carried out in lower dimensions \cite{CarlipNelson}, in 3+1 dimensions 
the resulting Hamiltonian is highly nonlocal, and fairly little progress has been made.} one imposes the
constraints (\ref{c2a})--(\ref{c3a}) as operator equations, most directly in the form
\begin{align}
{\hat{\mathscr{H}}}^i  \Psi[q] 
  &=  2i\hbar\left\{\,{}^{\scriptscriptstyle(3)}\nabla_j\frac{\delta\ }{\delta q_{ij}}\right\} \Psi[q] =0
 \label{db1}\\
{\hat{\mathscr{H}}} \Psi[q] &=  \left\{\frac{\hbar^2\kappa^2}{\sqrt{q}}
    G_{ijkl}\frac{\delta\ }{\delta q_{ij}}\frac{\delta\ }{\delta q_{kl}}
    - \frac{1}{2\kappa^2}\sqrt{q}\,({}^{\scriptscriptstyle(3)}\!R - 2\Lambda) \right\}\Psi[q] = 0  
\label{db2}
\end{align}
where $G_{ijkl} = q_{ik}q_{jl} + q_{il}q_{jk} - q_{ij}q_{kl}$ is the DeWitt metric 
on the space of metrics. 

Eqn.\ (\ref{db2}) is the original form of the famous Wheeler-DeWitt equation \cite{DeWitt1}. Other, not 
entirely equivalent, forms may appear in different quantization programs---in loop quantum gravity, for instance, 
the wave function $\Psi$ is a functional of a connection rather than the spatial metric---and the construction 
of an inner product on the space of solutions is a deep and largely unsolved problem \cite{Woodard}.  But it
seems likely that whatever the eventual final theory of quantum gravity, some echo of this equation
will remain.

We can now ask three questions:
\begin{enumerate}
\item Does a typical physically sensible wave function give significant probabilities for three-geometries 
that are ``foamy'' at the Planck scale, such as those described in \S\ref{initial}?
\item If it does, how does this structure evolve in time?  
\item Does canonical quantization allow topology change?
\end{enumerate}

We do not have an answer to any of these questions, but the answer to the first is plausibly ``yes.''  
By construction, the three-geometries of \S\ref{initial} solve the classical constraints, so it seems reasonable 
to expect the existence of corresponding coherent states in a quantum theory, satisfying the corresponding
quantum constraints.  Indeed, the only way to exclude a particular spatial metric $q_0$ from the quantum 
theory would be to require that $\Psi[q_0]=0$ for every wave function.  This seems a rather unnatural
restriction, and not one that arises in any obvious way from the formalism.

It is somewhat more plausible that certain classes of spatial metrics, while not excluded, might be
\emph{improbable} given a ``typical'' wave function.  This possibility is harder to address.  Most of what we know  
about the Wheeler-DeWitt equation comes from specializing to mini\-superspaces, that is, restricting
to certain classes of highly symmetric metrics, simply because more general cases are too hard 
\cite{Misner,Ryan}.  But such restrictions automatically exclude foamy spatial geometries.  One slightly 
more general investigation has considered the midisuperspace\footnote{A ``minisuperspace'' model freezes 
out all but a finite number of degrees of freedom; a ``midisuperspace'' model freezes out many degrees of 
freedom but retains an  infinite number.} of locally spherically symmetric metrics \cite{Carlipmini,Carlipmidib}.  Here,
geometrically (though not topologically) foamy three-geometries appear in the form of onion-like structures 
with nested layers of different geometries, some expanding and some contracting, joined by connected summation.  
The results so far indicate that foamy structures with rapidly varying geometries occur with fairly high probability, 
though there remain some unsettled questions about the proper probability measure.

The second question, that of time evolution, is harder, since it forces us to deal with the notorious 
``problem of time'' in quantum gravity \cite{Kuchar,Ishamb}.  Quantum gravity has no fixed background 
structure that can be used to measure the passage of time, and finding a general enough ``relational'' time 
is not easy.  For foamy spacetimes, in particular, traditional choices that depend on the expansion of space, 
like York's ``extrinsic time'' \cite{York} and the mean curvature flow \cite{Kleban}, cannot do the job 
\cite{Carlipmini}.  One alternative, used in \cite{Carlipmini},  is to make use of a ``cloud of clocks,'' a background 
of noninteracting test particles (``dust'') introduced to serve as a ``clock'' \cite{BrownKuchar}.  This is
not ideal, since the ``clocks'' back-react on the metric and affect the geometry.  But the backreaction can
be made small, and the resulting physical picture is close to the way we experimentally evaluate time
in the real world.   

But while the question of evolution is hard, it is also crucial.  Classically, the initial data of \S\ref{initial} leads 
to singularities: the ``necks'' of the connected sums typically contain marginally outer trapped surfaces, essentially
black hole horizons, and the standard singularity theorems apply \cite{Burkhartb}.  A fundamental question 
for the quantum theory is whether these singularities persist, or whether, in Wheeler's words, they are
``all the time and everywhere forming and disappearing, forming and disappearing\dots'' \cite{Wheeler4}.

The midisuperspace model of \cite{Carlipmini,Carlipmidib} has attempted to address this question.  By
introducing a matter reference system, a ``cloud of clocks,'' the model converts the Hamiltonian
constraint to a true physical Hamiltonian that can describe the evolution of ``foamy'' initial data.  The
Wheeler-DeWitt equation can be solved in a WKB approximation, and one can look for formally time-independent
stationary states.  Such states exist, and they frequently give high probabilities for ``foamy'' geometries.  One 
must be a bit careful about interpretation, though; even in ordinary quantum mechanics, we know that a formally 
time-independent WKB state can still describe a time-dependent process such as tunneling through a barrier.  
But the states found in \cite{Carlipmini} also have extremely small probability currents in the ``foamy''
regions of the space of geometries, and thus arguably describe 
nearly time-independent, self-reproducing spacetime foam, albeit in a still highly restricted setting.

Finally, though, even if canonical quantization allows complicated spatial geometry and topology, it is not
at all clear that it allows topology \emph{change}.  The wave function $\Psi[q]$ is most naturally defined
on superspace \cite{Wheelerx,Fischer}, the space of diffeomorphism classes of metrics on a fixed 
three-manifold $\Sigma$.  With such a definition, the topology of $\Sigma$ is fixed for all time.  There
may be a more general setting, though.  In two-dimensional Riemannian geometry, the equivalent of
superspace is the Teichm{\"u}ller space of metrics on a genus $g$ surface, and these spaces fit together
into a larger ``universal Teichm{\"u}ller space'' that includes all topologies \cite{Pekonen}.  A step towards
a three-dimension construction of a universal superspace can be found in \cite{Edwards}, though as far
as I know its connection to the Wheeler-DeWitt equation has not been explored.

\subsection{Lower dimensions}

As noted in \S\ref{other}, the geometry and topology of saddle points becomes much simpler in
three spacetime dimensions.  Moreover, the Einstein-Hilbert action can be rewritten as a Chern-Simons
action \cite{Achucarro,Witten1,Witten2}, a form in which the path integral (\ref{d2}) can often be computed
exactly.  This has allowed a number of calculations that may offer deeper insight into spacetime foam,
albeit in a somewhat unrealistic setting.
\begin{itemize}
\item{\bf Euclidean path integrals in three dimensions}\\[1ex]
We saw in \S\ref{other} that the saddle points of the Euclidean path integral in three dimensions are spaces
of constant curvature.  For $\Lambda<0$, these are the hyperbolic manifolds.  A partial sum over such
manifolds was evaluated, with one loop contributions, in \cite{Carliptopsum}, and was shown to diverge, 
although as explained in \S\ref{caveat}, one should interpret this with care.  By looking at a particular 
class of hyperbolic manifolds with a single boundary, it was shown in \cite{Carlipaction} that the Hartle-Hawking
wave function has infinite peaks, arising from an infinite number of distinct ``foamy'' topologies with the same
boundary metric.  (A higher dimensional version of this result  \cite{Carlippeaks} was mentioned  
in \S\ref{pathint}.)

For $\Lambda>0$, the Einstein manifolds are completely characterized.  Building on \cite{Carliptopsum},
the exact path integral was computed in \cite{Maloney} for one infinite class of such manifolds, the lens
spaces.  The sum fails to converge, even with zeta function regularization, although again this might 
change if more topologies are incorporated.  

\item{\bf Lorentzian path integrals in 2+1 dimensions}\\[1ex]
The relationship of (2+1)-dimensional gravity to Chern-Simons theory gives us some control over the
Lorentzian path integral as well, and there has been some work related to spacetime foam.  In
\cite{CarlipAlwis}, the partition function for an arbitrary number of wormholes was computed, in the
saddle point approximation for $\Lambda>0$ and with one loop corrections included for $\Lambda<0$.
The sum over wormholes was shown to diverge, and to not even be Borel summable.

The Lorentzian path integral also provides a formalism for computing amplitudes for spatial topology
change \cite{Witten2}, and thus for the ``production'' of the sort of spatial foam considered in \S\ref{initial}.
An explicit calculation can be found in \cite{CarlipCosgrove}, where it was argued that certain infrared
divergences may suppress topology-changing amplitudes, but this matter is not fully resolved.

In a somewhat different approach, a matrix model inspired by causal dynamical triangulations
(\S\ref{lattice}) can be used to obtain transition amplitudes between spatial slices \cite{AJLV}.   While 
the slices have a fixed $S^2$ topology at weak coupling, in the strong coupling regime they ``disintegrate''
into a foam of multiple components connected by wormholes.

\item{\bf Nonperturbative methods}\\[1ex]
Rather than computing an object such as the partition function for individual topologies and then 
summing, one can try to reorganize the calculation to include a sum over topologies from the start.
In two dimensions, matrix models do this: surfaces of varying topology are approximated by 
Feynman graphs of a suitable field theory, which then, in an appropriate limit, automatically
generates a sum over both geometries and topologies \cite{Gross}.  Boulatov has proposed a 
three-dimensional version of the same idea \cite{Boulatov}, in which the Feynman diagrams of a  
field theory form triangulated three-manifolds, which may have arbitrary topologies.  This model
has been modified to a form that can be shown to be Borel summable \cite{Freidel}.  While the result
not quite ``ordinary'' three-dimensional gravity, it offers a proof in principle that a sum over topologies
may make sense.  The four-dimensional version of this idea, group field theory, will be discussed
briefly in \S\ref{otherq}.

\item{\bf Two-dimensional gravity}\\[1ex]
The Einstein-Hilbert action is trivial in two dimensions, but various modifications lead to Einstein-like 
models of gravity.  The most famous is the Jackiw-Teitelboim model \cite{Jackiw,Teitelboim2}, but 
there are more general models of dilaton gravity \cite{dilaton}, which can be obtained from higher dimensional 
general relativity by dimensional reduction.  There is a vast literature on this subject, much of it involving sums 
over geometric and topological fluctuations.  In a sense, the whole Polyakov path integral approach to
string theory \cite{Polchinski2} can be thought of as a theory of two-dimensional conformal fields in a
background of fluctuating geometry and topology, complete with a dilaton gravity action for the
noncritical string.  I will not try to summarize this topic here, but will briefly mention two provocative
results: the use of matrix models to provide a nonperturbative definition of the theory \cite{Gross}, and
very recent work on two-dimensional gravity, the AdS/CFT correspondence, and the SYK model
\cite{Harlow,Shenker}.
\end{itemize}

\subsection{Other approaches \label{otherq}}
There are, of course, many other research programs aimed at to quantizing gravity, which may or 
may not exhibit spacetime  foam.  To a certain extent, this is a question of how broadly one defines ``foam.''  
In loop quantum gravity \cite{Ashtekar}, for instance, areas and volumes are quantized, with minimum 
values at the Planck scale.  Such a structure might be considered spacetime foam.  On the other hand, 
the spacing between allowed values is typically much smaller than Planck scale \cite{Agullo}.  In string 
theory, quantum production of virtual $D$-branes has been interpreted as a form of spacetime foam 
\cite{Ellis1,Ellis2}.  This proposal has not received a lot of theoretical attention, but there have been some
interesting explorations of possible observational implications, which I will return to in \S\ref{prop}.

There are a few settings and results that are worth mentioning explicitly:

\begin{itemize}
\item{\bf Spin foams and group field theory}\\[1ex]
A spin network is essentially a labeled graph, with edges labeled by representations of a group (usually
$\mathrm{SU}(2)$) and vertices labeled by group intertwiners.  Spin networks are states in loop quantum gravity%
---they determine mappings from the space of generalized connections to $\mathbb{C}$---and in fact they
form a  basis \cite{Ashtekar}.  A spin foam is the time evolution of a spin network \cite{Baez,Perez}.  As the 
name suggests, such an object looks foamy, with collection of surfaces (a two-complex) joining along edges 
in a soap-bubble-like  pattern.  This is not quite the same metaphor as Wheeler's, though, and it is not obvious 
whether spin foams necessarily imply spacetime foam, except in the very broad sense of having structure at the 
Planck scale.

There is, however,  one way in which spin foam models touch directly on the issue of spacetime
foam.  As noted earlier, sums over topologies tend to diverge.  Such divergences do not necessarily doom
the program---see \S\ref{caveat}---but they certainly complicate path integral treatments of topological fluctuations.  
Group field theories \cite{Oriti,Freidel1} are field theories whose Feynman diagrams are cell complexes dual to 
triangulated manifolds, with interactions that can be chosen to reproduce spin foam amplitudes.  A perturbative 
expansion of a group field theory thus reduces to the sum over spin foams that occurs in ordinary spin
foam models, but a nonperturbative evaluation naturally includes a sum over topologies.  The three-dimensional 
version due to Boulatov can be made Borel summable \cite{Freidel}, but it is not currently known whether this
extends to higher dimensions.  Still, these models offer one of very few ways we know that might make sense
of a sum over topologies.

\item{\bf Fractal spacetimes}\\[1ex]
In discrete approaches to the path integral, the integral seems to be dominated by very irregular,
nonsmooth configurations.  This is not so surprising: even in nonrelativistic quantum mechanics, the path
integral is dominated by paths that are nowhere differentiable \cite{Morette}.  Other signs of fundamentally
nonsmooth manifolds occur as well, including the appearance of dimensional reduction \cite{Carlipdim}
and minimum length \cite{Hossenfelder} in many approaches to quantum gravity.  These results suggest 
that perhaps the class of smooth manifolds discussed in \S\ref{construct} is too limiting.

Apart from the simplicial methods described in \S\ref{lattice} and a few proposals for small generalizations 
of manifold structure \cite{Hartleb,Schleich,Schleichb}, there has been fairly little work on this prospect
within conventional quantum gravity.  There has, however, been a research program aimed at understanding
properties of field theories in spaces with fractal dimensions \cite{Calcagni,Calcagni1,Calcagni2,Crane}, and
perhaps their implications for quantum gravity \cite{Calcagni3}.
 
\end{itemize}

\section{Fields in foamy spacetimes \label{fields}}\setcounter{footnote}{0}

The geometry of spacetime determines the gravitational field.  But spacetime is also the arena in which other
fields propagate, and it is possible that a foamy structure could influence that propagation.  This is not
entirely clear---the Planck scale is very small, and it is plausible that structures at that scale would wash
out over larger distances---but as possible signatures of quantum gravity, such effects have been quite
extensively investigated.

Broadly speaking, there are two ways one might study the effects of spacetime foam on particle behavior.
The first is to look at a particular model of foam, such as a Euclidean instanton approach (for instance,
\cite{Hawking_prop}),  to obtain detailed but model-dependent conclusions.  The second is to construct a 
broad phenomenological model (for instance, \cite{Garay,Ng}) to explore a range of possibilities.  In this section 
I will briefly describe both approaches, and then discuss the interesting possibility, again originally due to 
Wheeler \cite{Wheeler2}, that spacetime foam may soften or eliminate the divergences of quantum field theory.  

Quantum gravity phenomenology is a very large subject \cite{Amelino}, and I cannot hope to do it justice here.  
The reader should view this section merely as a taste of what's out there, and a hint of where to look 
for more.

\subsection{Propagation \label{prop}}

The analysis of quantum propagation in ``foamy'' spacetimes is limited both  by the scarcity of detailed models and
by the difficulty of calculation in settings in which explicit expressions for the metric are rarely known.  One early
attempt by Hawking, Page, and Pope \cite{Hawking_prop,Hawking_prop2} looked at the approximate behavior of
Greens functions on Riemannian manifolds formed as connected sums of $S^2\times S^2$ and $\mathbb{CP}^2$
``bubbles,'' with simple (non-Einstein) metrics.  At energies far enough below the Planck scale, the foamy structure
had little influence in S-matrices for spin-1/2 and spin-1 particles, but induced Planck-size $\lambda\varphi^4$ 
interactions, and possibly Planck-scale tachyonic masses,  for scalar fields.  This led the authors to argue that  
scalar particles such as the Higgs cannot be fundamental, though they  might occur as composite particles.   

In the same general spirit but with rather different details,
Friedman et al.\ analyzed fermion propagation on manifolds with nonorientable handles \cite{FPPZ}, and concluded
that fermions would acquire Planck-scale masses unless either the particles had a finite size (larger than that of typical
quantum fluctuations) or the path integral included a sum over spin structures and an arbitrary restriction to only 
CP-reversing handles.  Gravitational instantons, in particular K3 ``foam,'' could also generate fermion masses
through axial anomalies, but this effect appears to be strongly suppressed \cite{Hebeckerb}.

In a very different model, Ellis, Mavromatos, and Nanopoulos have considered propagation in a ``foam'' of D-branes,
$D$-dimensional string theory solitons \cite{Ellis1,Ellis2}.  They find that recoil off such solitons induces an energy
dependent vacuum refractive index, which might be astrophysically observable.   

As noted in \S\ref{passive}, there has also been some investigation of the effect of quantum stress energy fluctuations 
on the behavior of null and timelike geodesics.  Since Greens functions in a curved spacetime are determined by
geodesic distances, it may be possible to recast these results as predictions for field propagation, although some
work would be required to appropriately translate numerical results.  Note that the problem here is not to
understand the behavior of the fields producing the vacuum fluctuations---that is already under control---but rather to
analyze propagation of other fields in the geometry produced by those fluctuations.  Some preliminary calculations 
of the effect of purely gravitational fluctuations on the propagation of light have also recently been carried out in a 
symmetry-reduced simplicial path integral \cite{Jia}, and provide further evidence that such quantum effects
increase at small scales.

Given the difficulty of finding and working with realistic models of spacetime foam, though,  most of the work on quantum
field theory has been phenomenological.  That is, plausible effects of Planck scale fluctuations have been identified and
parametrized, with the goal of obtaining observational limits (see \S\ref{obs}).  For instance, scattering off quantum
foam is likely to be energy dependent, plausibly modifying dispersion relations for light.  If fluctuations of the metric over 
a proper distance $\ell$ are of the order
\begin{align}
\delta g\sim \left(\frac{\ell}{\ell_p}\right)^\alpha
\label{ea1}
\end{align}
(where $\ell_p$ is again the Planck length), the corresponding fluctuations of energy and momentum could
lead to modified dispersion relations of the form \cite{Ng1,Ng2}
\begin{align}
E^2 \sim p^2 \pm \epsilon\left(\frac{E}{E_p}\right)^\alpha
\label{ea2}
\end{align}
where $\epsilon$ is of order 1.  Similarly, for light with wavelength $\lambda$ one might expect random phase fluctuations 
on the order \cite{Lieu,Ng3}
\begin{align}
\delta\phi \sim 2\pi\delta\ell/\lambda \sim \frac{2\pi}{\lambda}\ell^{1-\alpha}\ell_p^\alpha .
\label{ea3}
\end{align}

Different choices of $\alpha$ correspond to different phenomenological models of spacetime foam.  Wheeler's
argument, described in \S\ref{Whee}, corresponds to $\alpha=1$.  The choice $\alpha = \frac{1}{2}$ has been
called ``holographic foam'' \cite{Ng4}, while $\alpha=\frac{2}{3}$ can be  characterized as a random walk
model \cite{Ng4,Amelino2,Diosi}.

Note that the dispersion relations (\ref{ea1}) are not Lorentz invariant.  This is not necessarily a problem, since
in an inhomogeneous spacetime the ``foam'' provides a background structure.  But one might expect that a quantum 
treatment of such fluctuations would restore at least statistical Lorentz invariance.  As far as I know, this issue has
not been explored directly in the context of spacetime foam.  However, Hossenfelder has looked
at a similar problem in which ``foam'' is replaced by pointlike defects, and has shown that statistical Lorentz invariance
can, in fact, be recovered \cite{Hossenfelder2}.

\subsection{Decoherence and unitarity \label{decoh}}

The idea that spacetime foam could lead to a failure of unitarity in quantum field theory 
dates back to work by Hawking \cite{Hawking_un,Hawking_un2}, who argued that virtual black holes could 
form coherently and then evaporate thermally.  The 
plausibility of this argument depends on the answer to the ``information loss problem,'' the question of whether the 
formation and subsequent evaporation of black holes is unitary.  I think it is fair to say that most physicists currently 
working on quantum gravity expect that the process is ultimately unitary, but the question is still hotly debated 
\cite{Marolf_un,Wald_un}.

A similar argument for loss of coherence can be made for topology changes in which ``baby universes'' split
off \cite{Lav}.  But even if black hole evaporation and similar processes are unitary,  it remains possible that spacetime 
foam could lead to \emph{effective} decoherence and nonunitarity.  Fields in a foamy spacetime may behave as open 
systems, with the Planck scale fluctuations of the spacetime acting as an environment; if the fields are observed and 
the spacetime is not, the result can be a loss of coherence \cite{Mav}.  Moreover, virtual handles and similar nontrivial 
fluctuations of topology can lead to effective nonlocal interactions, which can perhaps act as a thermal bath 
\cite{Garay,Garay2}.   It has been argued that such decoherence could induce CPT violation, possibly at an observable 
level \cite{Sarkar,Mav}.

Even this much is not certain, however.  Coleman and others have argued that wormholes (and plausibly other 
topological fluctuations) do not, in fact, lead to nonlocal interactions, but can rather be described by local operators 
that have an interpretation as new constants of nature \cite{Coleman,Coleman2,Gidd,Preskill}.  This argument---or 
at least its applicability  to our Universe---has been questioned (see, e.g., \cite{Polchinski}), but there has been a recent 
revival of interest in the context of black hole unitarity \cite{Marolf2}.  I will return to this idea in a bit more detail
in \S\ref{CC}; for a recent review, see \cite{Hebecker}.

Of course, there are other proposals in which quantum gravitational effects can lead to decoherence, especially 
if, as Penrose has suggested, quantum gravity requires fundamental modifications of quantum mechanics 
\cite{Penrose}.  In particular, spontaneous collapse models have been proposed in which the collapse probability 
increases in regions of high curvature \cite{Modak}, as might occur in ``foamy'' spacetimes.

\subsection{Divergences}

In his first proposal of spacetime foam, Wheeler suggested that 
\begin{quotation}
\dots it is essential to allow for fluctuations in the metric and gravitational interactions in any proper treatment 
of the compensation problem---the problem of compensation of ``infinite'' energies that is so central to the 
physics of fields and particles \cite{Wheeler2}.
\end{quotation}
This hope---that quantum gravity might somehow eliminate the divergences of quantum field theory,
perhaps by providing a Planck scale cutoff---has persisted as an impetus for studying spacetime foam. 
It gained plausibility from the discovery, initially by DeWitt \cite{DeWitt3} and then extended by
others \cite{Khrip,Salam,Isham2}, that certain infinite sums of Feynman diagrams containing gravitons are 
finite even though the individual diagrams diverge.  Unfortunately, this phenomenon has only been shown to 
hold for certain special classes of diagrams.  It has been argued that to obtain a general result, one must 
systematically reorganize the perturbation expansion to allow cancellations between different orders
\cite{Woodard2}, but at least for now there is no conclusive demonstration.

In the absence of exact results, a number of physicists have turned to more phenomenological models.
It was noted long ago \cite{Landau,Pauli,Deser} that quantum gravitational fluctuations at the Planck scale
might ``smear'' the light cone, potentially eliminating the source of divergences.  A number of relatively
ad hoc smearings have investigated \cite{Modesto,Padmanabhan,Abel,Kan}, and do indeed cut off 
divergences;  Ohanian has argued that this behavior is independent of the details of the smearing \cite{Ohanian}.  
In a variation on this idea, Casadio has proposed a modification in which each virtual particle in a Feynman 
diagram moves in a background determined by the other particles in the diagram \cite{Casadio}.  More
formally, Haba has analyzed the behavior of a quantum field theory in a background metric undergoing
random scale invariant fluctuations, again leading to better behaved propagators \cite{Haba}, and Ford
has considered propagation in a background of gravitons in a squeezed state, which again removes
propagator singularities \cite{Ford_div}

In a more explicitly ``foamy'' construction, Crane and Smolin have considered the effect of a gas of
virtual black holes on field propagation \cite{Crane,Crane2}, arguing that their presence alters
the spectrum of perturbations in a way that imposes a high energy cutoff on the density of states.
Interestingly, this mechanism also reduced the effective short distance dimension of spacetime.
A similar ``dimensional reduction'' occurs in a number of other approaches to quantum gravity
\cite{Carlipdim}; it is not known whether the phenomenon is generically connected to spacetime
foam

\section{Observational tests \label{obs}}\setcounter{footnote}{0}

If the standard picture of spacetime foam is right, quantum fluctuations of spacetime only become
large near the Planck scale.  Even our most energetic probes  have wavelengths far 
larger than $\ell_p$, and thus cannot resolve individual Planck scale events.  It remains
possible, though, that the cumulative effect of many Planck scale fluctuations might be observable.
This is not unprecedented in physics: Brownian motion, for instance, demonstrated the existence
of molecules well before we could directly probe molecular scales.

The search for observational effects of spacetime foam is part of the larger field of quantum gravity
phenomenology.  This is too large a subject to discuss in detail here; for some recent reviews, see
\cite{Amelino,Ng_rev,Karpacs,Addazi}.  Instead, this section will describe some of the conceptual difficulties
in formulating observational questions, and then very briefly list some of the proposals for, perhaps,
detecting spacetime foam.

\subsection{The problem of averaging}

In practice, any observations we can make will probe a region much larger than Planck size.  Classically,
we see spacetime averages; quantum mechanically, the observables relevant to our experiments must
similarly be averaged.   But in a curved spacetime, such averages are difficult to define:
\begin{itemize}
\item[--] Components of tensors depend on a choice of basis, which can vary from point to point.  To
define averages, one must compare bases, for instance by parallel transporting back to a basepoint
(see, for example, \cite{Hoogen}).  But even apart from its technical difficulty,  such a procedure only
makes sense if one can define a unique path to the basepoint.  In a ``foamy''  region with very high 
local curvature, the existence of caustics will restrict such a procedure to very small regions.
\item[--] Even to average a scalar, one must define the region over which one is averaging, a process
that typically depends on the geometry.  One might, for instance, pick a time slice and average over 
geodesic balls of radius $\varepsilon$.  But both the definition of the slice (say as a slice of constant
mean curvature) and the geometry of the geodesic balls will depend on the metric, making it  
difficult to compare results among different manifolds, or even different locations in the same manifold.
\end{itemize}

These problems have mainly been studied in the context of cosmology (see, for instance, \cite{Buchert}),
where they have generated quite a bit of controversy \cite{Green,Buchert2}.  But they are clearly relevant 
to the understanding of macroscopic effects of spacetime foam---both for averaging over
geometries and, for a fixed inhomogeneous geometry, for comparing different locations and times.  In
principle, a full quantum theory of gravity could give us answers, but even there, there are deep
conceptual difficulties in defining local or quasilocal observables \cite{Torre,Giddings,Giddings2}.

\subsection{Some proposed tests}

These difficulties have not, of course, prevented physicists from looking for observational effects of
spacetime foam.  In particular, phenomenological models of the sort described in \S\ref{fields} may make
plausible predictions even in the absence of detailed quantitative calculations.  Here I list some of the possible%
---although by no means certain---effects.  Note that the references here are quite incomplete; they are 
meant only to give the reader a starting point.

\begin{itemize}
\item{\bf Lorentz breaking}\\[1ex]
As discussed in \S\ref{prop}, spacetime foam could provide a preferred reference frame, allowing for a
violation of Lorentz invariance.  This is one of a number of motivations for searches for Lorentz breaking,
with observations ranging from laboratory to astrophysical scales.  This is a very active area of research; 
for reviews, see \cite{Mattingly,Liberati,Heros}.

Most observational searches have looked for what is sometimes called ``systematic'' breaking of Lorentz
invariance, breaking in which there is a global preferred frame.  Modified dispersion relations such as
(\ref{ea2}), for instance, assume a global choice of time to distinguish momentum from energy.  It is
perhaps more plausible for spacetime foam to lead to stochastic, local violations, or ``nonsystematic''
breaking.  Such effects are harder to observe, of course, but there have been some interesting limits
\cite{Hossenfelder2,Basu,Vas}.  

\item{\bf Decoherence and CPT breaking}\\[1ex]
As described in \S\ref{decoh}, some phenomenological models of spacetime foam predict decoherence and
a breakdown of unitarity, either at a fundamental level (for example, if black hole formation and evaporation
is nonunitary) or an effective level (if correlations with quantum fluctuations are neglected).  
This decoherence can pick out a preferred direction of time, leading to CPT violation.  Both decoherence
(for instance, in neutrino or $K$--${\bar K}$ oscillation) and CPT violation might be detectable in elementary
particle experiments \cite{Sarkar,Mav,Carrasco,Stuttard}.

\item{\bf Blurring/incoherence of starlight}\\[1ex]
Spacetime foam may also fairly generically lead to a breakdown of the \emph{classical} coherence of light,
as different parts of a wave front pass through different Planck scale geometries \cite{Lieu,Ng3}.  Such a
loss of coherence would lead to the disappearance of diffraction patterns---Airy rings---from extragalactic
sources, as well as angular broadening and blurring of distant point sources.  The observed absence of such
effects has been used to rule out certain phenomenological models of spacetime foam \cite{Lieu,Ragazzoni,%
Tamburini,Steinbring,Perlman}, although other models suggest that the effects may still be unobservably small 
\cite{Mazi}.

\item{\bf Interferometer noise}\\[1ex]
If the effects of spacetime foam are large enough, decoherence may also be an observable source of noise
in interferometers \cite{Amelino_gw,Ng_int,Hogan}.  In some ``holographic'' models of spacetime foam, the
relevant quantum fluctuations have macroscopic correlations (see, for instance, \cite{VerlindeZurek}). 
These have allowed experimental searches and placed limits on particular phenomenological models 
\cite{Holo,Holo2,Vermeulen}.

\item{\bf Broadening of spectral lines}\\[1ex]
If photons can lose energy through interactions with spacetime foam, one effect should be a broadening of
spectral lines.  Measurements of absorption lines in extragalactic gas clouds have been used to restrict some
phenomenological models \cite{Thompson,Cooke}.

\item{\bf Cosmology}\\[1ex]
Spectral broadening from spacetime foam will also affect thermal spectra, and in particular the extremely
well measured spectrum of the Cosmic Microwave Background \cite{Stefano}.  The effect is strong enough
to rule out a certain class of phenomenological models.

\item{\bf Axions}\\[1ex]
If axions exist, their shift symmetry can be broken by spacetime fluctuations, in particular by wormholes 
\cite{Rey}.  The resulting potential can influence axion-driven inflation, as well as phenomena such as black
hole superradiance and axion models of dark matter.  For a review, see \cite{Hebecker}.

 \item{\bf Particle propagation}\\[1ex]
As noted in \S\ref{prop}, a particular approach to spacetime foam, a partial sum over Euclidean instantons,
seems to predict Planck scale masses and interactions for the Higgs boson and fermions 
\cite{Hawking_prop,Hawking_prop2,FPPZ}, and is thus ruled out by observation.  But this example
also illustrates one of the pitfalls of spacetime foam phenomenology.  The calculations leading to
this conclusion assumed that the Higgs and fermions were noncomposite and pointlike down to the
Planck scale.  While this is consistent with our present knowledge of quantum field theory, it requires
extrapolations to scales far beyond what can be probed by current observations.  So while the results
of \cite{Hawking_prop,Hawking_prop2,FPPZ} can be taken to rule out \emph{something}, what is
excluded is more elaborate than simply a particular model of spacetime foam. 

\item{\bf Minimum length}\\[1ex]
The ideas of spacetime foam and ``minimum length'' are logically distinct, although both could arise from 
Planck scale quantum fluctuations.  Still, one might argue that certain types of spacetime foam could
imply a minimum length or a ``generalized uncertainty principle'' \cite{Adler} by directly affecting
the measuring process; see \cite{Scardigli} for an early example.  There is an enormous literature on 
possible observational tests of a minimum length, which I will not try to summarize here; see 
\cite{Hossenfelder} for a recent review.
\end{itemize}

\section{The cosmological constant \label{CC}}\setcounter{footnote}{0}

Let us finally turn to one phenomenon for which it has been proposed that spacetime foam might have a large
macroscopic effect, the ``cosmological constant problem.''

Gravity is universal: it couples to all forms of energy with equal strength.  This principle is a cornerstone of
general relativity---it's the reason gravity can be viewed as geometry---and it has been tested to accuracies of
a part in $10^{15}$ \cite{MICRO}.  The problem is that ``all forms of energy'' should include the energy of
vacuum fluctuations, which should manifest as a cosmological constant $\Lambda$.  There are arguments 
over the nature of this contribution \cite{Holland}, but conventional methods give an answer that seems to disagree 
with observation by a factor of at least $10^{58}$, and perhaps as much as $10^{122}$ \cite{Martin,Weinberg}.%
\footnote{The commonly quoted factor of $10^{122}$ comes from a non-Lorentz-invariant
cutoff.  If one uses an invariant method such as dimensional regularization or Pauli-Villars, a field of mass $m$
contributes on the order of $m^4\times\hbox{\it logs}$.  If the top quark is the heaviest particle, this gives a
factor of about $10^{58}$; if new particles continue up to the Planck mass, one again obtains $10^{122}$.} 

In the early 1980s, Baum \cite{Baum} and Hawking \cite{Hawking_cc} argued that the Euclidean path integral 
in the presence of a positive cosmological constant is dominated by the contribution of the four-sphere,
giving
\begin{align}
Z \sim e^{3\pi/{\ell_p^2\Lambda}}  .
\label{f1}
\end{align}
There are debates about the sign in the exponent (for instance, \cite{Turok,Marolf_eucl}),
but if (\ref{f1}) is correct and $\Lambda$ is positive, it suggests a very high probability for a very small 
cosmological constant.

This is not quite right, of course---in standard general relativity, $\Lambda$ is a fixed parameter, not a variable
in a probability amplitude.  Coleman has argued that spacetime foam can change this \cite{Coleman,Gidd}.
(For a nice description, see section VIII of \cite{Weinberg}.)  At scales that are large compared
to the scale of quantum fluctuations of spacetime, wormholes and ``baby universes'' should have effective
descriptions in terms of local operators, changing the gravitational action to
\begin{align}
I \rightarrow I + \sum_i(a_i + a^\dagger_i)\int\!d^4x\sqrt{|g|}\mathcal{O}_i
\label{f2}
\end{align}
where $a_i$ and $a^\dagger_i$ are annihilation and creation operators for fluctuations of type $i$ and
the $\mathcal{O}_i$ are some set of local operators.  In an eigenstate $|\alpha_i\rangle$ of $a_i + a^\dagger_i$,
the effect of this change is to shift the coupling constants in the action for the operators $\mathcal{O}_i$, 
including in particular the cosmological constant.  In a superposition of states $|\alpha_i\rangle$ 
with different eigenvalues $\alpha_i$, a path integral of the form (\ref{f1}) would overwhelmingly favor states 
in which $\Lambda$ was very near zero.  

As noted earlier, this argument is controversial \cite{Polchinski,Fischler,Colemanc,Hebecker}.
There is a danger that it would lead to ``giant wormholes''; the role of configurations that are only 
``nearly'' saddle points is unclear; and the results depend crucially on the controversial sign in the exponent 
(\ref{f1}).  There are also arguments that the Coleman mechanism is inconsistent with the AdS/CFT
correspondence of string theory \cite{Nima}, suggesting that certain topologies might be
excluded in a complete quantum theory of gravity.  Still, though, at the very least the argument shows that
spacetime foam could have a large macroscopic effect.

Vacuum fluctuations of bosonic and fermionic fields contribute to the cosmological constant with opposite 
signs, and quantum field theoretical considerations do not tell us whether $\Lambda$ should be positive
or negative.  If $\Lambda<0$, a completely different set of saddle points contribute to the Euclidean path
integral.  As discussed in \S\ref{pathint}, the sum over topologies (\ref{d6}) fails to converge: any process
that would normally push $\Lambda$ more negative instead drives the production of new topologies.
It has been suggested that this mechanism could explain the absence of a large negative cosmological
constant \cite{CarlipEucl}.  If, on the other hand, one believes that asymptotically anti-de Sitter spaces should 
be allowed in quantum gravity \cite{AdSCFT}, this result may alternatively be taken as another demonstration
that certain categories of topologies should not be summed in the path integral \cite{Farey}.

A somewhat different approach to spacetime foam and the cosmological constant is based on canonical
quantum gravity.  Refs.\ \cite{Carlip0,Carlip_essay,Carlipmini,Carlipmidib} start  with a naive question: 
how would we know if $\Lambda$ 
really were large?  In a homogeneous universe, of course, this is easy: a cosmological constant almost always
leads to an exponential expansion or contraction of space \cite{Waldcc}.  But the cosmological constant we
are interested in comes from Planck scale fluctuations, and on that scale the universe is certainly not 
homogeneous.   In Wheeler's original picture of spacetime foam, large fluctuations at the Planck scale 
average out to be nearly invisible at macroscopic scales.  Could that happen here?

It is useful to split this question into two parts: is there initial data that ``hides'' a large cosmological 
constant, and, if so, is that behavior preserved under time evolution?  Ref.\ \cite{Carlip0} gives an affirmative 
answer to the first question.  As discussed in \S\ref{canon}, the work of Chrusciel et al.\ \cite{Chrusciel,Chruscielb}
can be used to construct an infinite set of rapidly fluctuating ``foamy'' initial data for which the macroscopic 
expansion averages to zero even with a large cosmological constant.    Indeed, if space is dominated at the 
Planck scale by quantum fluctuations with no preferred direction of time, such behavior is probably generic.

The evolution problem is harder.  Naively one might expect that expanding regions grow and contracting regions 
to shrink, leading  to an eventual domination of expansion.   But if quantum fluctuations continue, expanding regions 
might themselves full up with ``foam,'' sustaining the initial low expansion structure.  

The question is, in any case, fundamentally 
one for quantum gravity, since the relevant initial data evolves classically to singularities that must somehow 
be avoided \cite{Burkhartb}.  While this is too hard a problem for a full quantum treatment, \cite{Carlipmini}
analyzed a spherically symmetric midisuperspace model that incorporated many of the features of the
full problem.  The resulting Wheeler-DeWitt equation can be solved in a WKB approximation, and the wave
functions have interesting properties:  while they have some support on purely expanding and purely 
contracting geometries, the most probable configurations are ``foamy'' geometries with a mix of expanding and
contracting regions, yielding very small average expansion.  Moreover, these ``foamy'' regions are long-lasting%
---the probability current for moving out of such regions is tiny---lending support to the idea that small 
macroscopic expansion may persist even with a large $\Lambda$.

A  somewhat similar mechanism may come into play even if $\Lambda$ is set to zero by a fine tuned choice of 
renormalization \cite{Wang1}.   In this case, as described in \S\ref{passive}, one still expects enormous
\emph{fluctuations} in the vacuum energy, leading again to a very highly inhomogeneous spacetime.  Unruh
and Wang argue that these fluctuations average to almost zero, but that a parametric resonance produces an
overall expansion rate that is nonzero but exponentially smaller than the naive expectation.  In a later paper
\cite{Wang2}, they suggest that a large negative bare cosmological constant could combine with a positive 
contribution from spacetime foam to produce a ``microcyclic'' structure to the Universe, with Planck sized regions
independently undergoing oscillations but with a net overall expansion.

\section{Conclusion}

Spacetime foam is a vague and speculative idea, and it is tempting to postpone it until we have a full quantum
theory of gravity.  But even in the absence of such a theory, the notion of large quantum fluctuations at the
Planck scale may have interesting implications, some of them perhaps even observationally testable.  And
if we are lucky, the study of such implications might also help us in the larger problem of quantizing gravity.

As we have seen, spacetime foam could take many different forms.  Quantum fluctuations might affect
topology, or just geometry.  They might affect four-geometry, or spatial three-geometry, or the more abstract
Euclidean geometry of path integral instantons.  They might lead to a wide range of possible observational
effects, but alternatively might be too small to see at the scales we can observe.  

In this review, I have tried to give a broad overview of what is known, and speculated, about Wheeler's
proposal.  I have undoubtedly missed a lot, but I hope I have provoked some thought.  I will finish, as I
began, with a provocative---perhaps right, perhaps wrong---quote from Wheeler \cite{Wheeler4}:
\begin{quote}
The track of the particle looks impressive in its passage through a Wilson chamber.  The white cloud,
too, looks impressive in the transparent sky.  However, the proper starting point in dealing with physics
in the one case is the sky, not the cloud; and we are free to believe that the proper starting point in
the other case is the physics of the vacuum, not the physics of the particle.  To adopt this perspective
does not yield any sudden illumination about either particles or the vacuum, but does at least suggest
that no theory of particles that deals only with particles will ever explain particles.
\end{quote}

\newpage
 \begin{flushleft}
\large\bf Acknowledgments
\end{flushleft}

I would like to thank Renate Loll for some detailed comments on an earlier draft of this paper.
 This work was supported in part by Department of Energy grant DE-FG02-91ER40674.

 \end{document}